\documentclass[english,pre]{revtex4}
\usepackage[]{fontenc}
\usepackage[latin1]{inputenc}
\usepackage{graphicx}

\makeatletter

\newcommand{\noun}[1]{\textsc{#1}}

\providecommand{\tabularnewline}{\\}

\usepackage{bm}

\usepackage{babel}
\makeatother
\begin{document}

\title{Parametric excitation and chaos through dust-charge fluctuation in
a dusty plasma}

\author{Madhurjya P Bora and Dipak Sarmah}

\affiliation{Physics Department, Gauhati University, Guwahati, India.}

\email{mpbora@yahoo.com}

\begin{abstract}
We consider a van der Pol-Mathieu (vdPM) equation with parametric
forcing, which arises in a simplified model of duty plasma with dust-charge
fluctuation \cite{saitou}. We make a detailed numerical investigation
and show that the system can be driven to chaos either through a period
doubling cascade or though a subcritical pitchfork bifurcation over
an wide range of parameter space. We also discuss the frequency entrainment
or frequency-locked phase of the dust-charge fluctuation dynamics
and show that the system exhibits 2:1 parametric resonance away from
the chaotic regime.
\end{abstract}
\maketitle

\section{Introduction}

The subject of parametric excitation can be traced back to Faraday
in 1831 \cite{faraday}, when he observed that surface waves in a
fluid-filled cylinder under vertical excitation exhibited twice the
period of the excitation itself. The most simplified version of parametric
excitation was given by Mathieu in 1868 \cite{mathieu} related to
the vibrations of an elliptical membrane, which now has become a model
equation for response of many systems to sinusoidal parametric excitation.
The simplest Mathieu equation can be stated as \cite{nayfeh},\begin{equation}
\ddot{x}+(\delta+\epsilon\,\cos t)x=0,\label{eq:mathieu}\end{equation}
with $\delta$ and $\epsilon$ as constants. There have been extensive
investigations of parametric excitation and resonance related Mathieu
equation by several authors \cite{rand1,rand2,rand3}. One of the
very common examples of parametric forcing modeled by the nonlinear
Mathieu equation is the forced and unforced inverted pendulum \cite{kim}.

In this work, we have studied the parametric excitation and resonance
of the van der Pol-Mathieu (vdPM) equation, which arises in a simplified
model of dusty plasma with dust-charge fluctuation. Dusty plasmas
are characterized by presence of massive dust (impurities) particles
embedded in an electron-ion plasma \cite{shukla}. Immersed in the
plasma, the dust particles acquire charges by collecting the electrons
and ions on their surfaces, which are mostly negative. However, the
charge on a dust particle is never a constant and varies temporally.
Thus, along with other their dynamical properties, the charge of the
dust particles becomes a dynamic variable, which can severely modify
the plasma properties. The presence of dust particles in a plasma
may modify the dynamics of the plasma in many ways, of which the most
prominent is the appearance of low frequency dust-acoustic waves \cite{shukla}.
The dust-charge fluctuation is known to damp the acoustic waves in
a dusty plasma \cite{jana,shukla}. Recently, Momeni, Kourakis, and
Shukla \cite{momeni} has studied a simplified model of nonlinear
dust-charge fluctuation based on a vdPM equation, where they have
discussed the stability regions of the vdPM equation and have shown
the existence of stable and unstable periodic orbits in different
parameter space. This vdPM equation is originally derived by Saitou
and Honzawa \cite{saitou}, where they have shown that in a very restricted
region of parameter space, the system exhibit chaotic behavior. They
argue that the oscillations leading to chaos basically stems out of
the balancing between the van der Pol (vdP) and Mathieu-like terms.

We report, in this paper, a detailed investigation of the vdPM equation
for dust-charge fluctuation and show that the vdPM equation proposed
by Saitou \cite{saitou} can exhibit chaotic behavior over an wide
range of relevant parameter space, which is, in many instances, preceded
by period doubling cascades. We have found that both period doubling
and pitchfork bifurcations take place as one varies the bifurcation
parameters, both leading to chaos. The range of parameters for a chaotic
regime comes out to be not as restrictive as pointed out by Saitou
\cite{saitou}. We have discussed the stability and bifurcation of
the nonlinear system and found that the system can be completely deterministic
in between chaotic regimes. The paper is organized as follows. In
Section II, we formulate the nonlinear dust-charge fluctuation model
yielding the vdPM equation and discuss about the parametric forcing.
In Section III, we have discussed about frequency entrainment (frequency-locked
phase) of the nonlinear oscillator, driven by the parametric forcing
term. We have shown that far away from the chaotic regime, the system
can be driven by a 2:1 parametric resonance leading to a stable limit
cycle. Away from the resonance, it displays quasi-periodic behavior.
We discuss the stability and bifurcation of the periodic orbit of
the nonlinear system in Section IV with the help of Floquet stability
theory and show that the instability of a limit cycle may manifest
through a period doubling bifurcation. In Section V, we explore the
chaotic regime of the vdPM equation, where we show that the system
can be driven to chaos over an wide range of parameters. We have found
that the route to chaos may be through a period doubling cascade.
We draw the conclusions in Section VI.

\section{Dust-charge fluctuation model}

We consider an unmagnetized collisionless plasma consisting of electrons,
ions, and massive dust particles, which become charged by acquiring
charged particles (ion or electrons) on their surfaces. The subject
of dust-charge fluctuation in a dusty plasma is a well studied process
which, primarily, has a damping effect on the acoustic waves \cite{jana}.
Here, we assume that the number densities of the charged particles
are considerably larger than that of the dust particles, so that the
effect of dust-charge fluctuation on the dynamics of the electrons
and ions is negligible and charge neutrality is always satisfied \cite{saitou}.
So, the charge on the dust grains $q(t)$ becomes a time dependent
function.

Assuming the equilibrium (unperturbed) state to be static, we can
write the nonlinear continuity, momentum balance, and the Poisson
equation for dust-charge fluctuation as \cite{saitou,momeni},\begin{eqnarray}
\frac{\partial n}{\partial t}+n_{0}\nabla\cdot\bm v & = & \alpha n-\frac{1}{3}\beta n^{3},\label{eq:continuity}\\
m_{d}\frac{d\bm v}{dt} & = & q\bm E,\label{eq:momentum}\\
\varepsilon_{0}\nabla\cdot\bm E & = & qn,\label{eq:poisson}\end{eqnarray}
where $m_{d},n_{0},\varepsilon_{0}$ are dust mass, equilibrium dust
number density, and permittivity of free space. The variables $n,\bm v,\bm E$
are perturbed dust density, velocity, and electric field. In the first
equation, Eq.(\ref{eq:continuity}), the terms on the right hand side
denote the rate of production and loss of charged dust grains, where
$\alpha$ and $\beta$ are constants of proportionality. In writing
these terms, we have assumed that the production rate of charged dust
particles is proportional to the dust density. The cubic loss term
appears mainly due to the loss of dust grains through a three-body
recombination process \cite{saitou}. In the momentum equation, Eq.(\ref{eq:momentum}),
we have assumed the dust particles to be cold which basically eliminates
any variant of dust-acoustic waves. In writing these equations, we
have assumed that the average dust velocity, $\bm v$, is fairly uniform
in space and its spatial gradient is considerably smaller so that
the convective derivative term in the momentum equation, namely, the
term $(\bm v\cdot\nabla)\bm v$ can be neglected. Further, we have
approximated the term $\nabla\cdot(n\bm v)$ with $n_{0}\nabla\cdot\bm v$,
assuming a uniform distribution of the charged dust particles in space
($\nabla n\approx0$) \cite{momeni}. We assume the dust-charge $q(t)$
to be changing harmonically with time with a frequency $\nu$ and
use the ansatz \cite{saitou,momeni},\begin{equation}
q(t)=q_{0}(1-\epsilon\lambda\,\cos\nu t)^{1/2},\label{eq:charge}\end{equation}
where the term $(\epsilon\lambda)$ denotes the strength of charge
fluctuation. Note that, in principle, there is no need to restrict
the value of the term $(\epsilon\lambda)$ to a smaller value, which
gives us freedom to explore an wider parameter space of $\alpha$-$\epsilon$.

Without loss of generality, we consider only one dimension, $z$ and
write Eqs.(\ref{eq:continuity}-\ref{eq:poisson}) as,\begin{eqnarray}
\frac{\partial n}{\partial t}+n_{0}\frac{\partial v_{z}}{\partial z} & = & \alpha n-\frac{1}{3}\beta n^{3},\label{eq:oned_continuity}\\
m_{d}\frac{\partial v_{z}}{\partial t} & = & qE_{z},\label{eq:oned_momentum}\\
\varepsilon_{0}\frac{\partial E_{z}}{\partial z} & = & qn.\label{eq:oned_poisson}\end{eqnarray}
By taking a $z$-derivative of Eq.(\ref{eq:oned_momentum}), we can
eliminate the terms involving perturbed velocity and electric field
using Eqs.(\ref{eq:oned_poisson}) and (\ref{eq:oned_continuity})
to get a coupled differential equation in perturbed density,\begin{equation}
\frac{d^{2}n}{dt^{2}}-(\alpha-\beta n^{2})\frac{dn}{dt}+n\omega_{d}^{2}(1-\epsilon\lambda\,\cos\nu t)=0,\label{eq:final_dusteq}\end{equation}
where $\omega_{d}=(n_{0}q_{0}^{2}/m_{d}\varepsilon_{0})^{1/2}$ is
the plasma frequency corresponding to the dust particles. The above
equation, Eq.(\ref{eq:final_dusteq}) can be classified as \emph{van
der Pol-Mathieu} (vdPM) equation \cite{saitou}, owing to the nonlinear
term $(\alpha-\beta n^{2})$ which is like a van der Pol (vdP) term
\cite{nayfeh} and parametric forcing term $(1-\epsilon\lambda\,\cos\nu t)$
which like the parametric term of a classical Mathieu equation \cite{nayfeh}.

\subsection{Parametric forcing}

As is well known from the theory of classical Mathieu equation, the
parametric forcing term in Eq.(\ref{eq:final_dusteq}) makes the dynamics
of the dust-charge fluctuation prone to chaos \cite{armbruster,blumel,jeong,kim}.
As the vdP equation has a stable limit cycle, we can see that Eq.(\ref{eq:final_dusteq})
should show vdP-type behavior for large $\alpha$. However the parametric
forcing term may still drive the system unstable. In all probability,
we expect the onset of chaotic behavior as $\epsilon$ increases,
which should be more pronounced when $\alpha\ll1$.

In absence of the parametric forcing term, the stable limit cycle
of the vdP equation has frequency of 1. Another well known result
from the analysis of Mathieu equation \cite{nayfeh} is that the origin
becomes unstable when the parametric forcing frequency $\nu$ is close
to twice the frequency of the unforced oscillator. So, when both the
vdP and Mathieu terms are present, as in Eq.(\ref{eq:final_dusteq}),
we expect a frequency entrainment at 2:1 \cite{pandey} and the system
represented by Eq.(\ref{eq:final_dusteq}) must exhibit some sort
of quasi-periodic and frequency-locked (entrainment) behaviors in
the parameter space of $\epsilon$-$\alpha$ before it can be driven
to chaos. The route to chaos should be through a series of quasi-periodic
regime or through a period-doubling cascade rather than the other
universal route i.e. through intermittancy \cite{strogatz}.

\section{Frequency entrainment}

Entrainment dynamics plays an important part in design engineering
and many other dynamical systems \cite{zalal}. Recently, frequency
entrainment is shown to exist in nonautonoums chaotic oscillators
\cite{bove}. In this work, we consider the possible entrainment by
the parametric term, which can lead to quasi-periodicity and finally
to chaos. We consider the following dynamical equation for this dust-charge
dynamics,\begin{equation}
\ddot{x}-(\alpha-\beta x^{2})\dot{x}+\omega_{d}^{2}x(1-\epsilon\lambda\,\cos\nu t)=0,\label{eq:dusteq}\end{equation}
where we have replaced the dust density $n$ by the variable $x$.
In order to facilitate the multiple scales in the problem and entrainment,
we assume that $\alpha\sim\beta=\delta\ll1$, a small number. As the
entrainment is possible only when the system as far away from the
chaotic regime, when $\epsilon$ is small, we assume that $\epsilon<1$.
We further re-scale the time by $t\rightarrow\omega_{d}t$ and write
Eq.(\ref{eq:dusteq}) as\begin{equation}
\ddot{x}-\mu\epsilon(1-x^{2})\dot{x}+x(1-\epsilon\lambda\,\cos2\omega t)=0,\label{eq:entrain}\end{equation}
where $\mu\epsilon=\delta/\omega_{d}$. The strength of the parametric
forcing term is given by $\epsilon\lambda$ with $\epsilon<1$ and
$\nu=2\omega$. We expect that the parametric forcing should result
in a 2:1 subharmonic resonance, when the parametric frequency $\nu$
is close to 2 or $\omega\sim1$. Note that in absence of the parametric
forcing term ($\epsilon\lambda=0$), the natural frequency of the
oscillator is unity.

We now introduce two different time-scales \cite{strogatz,pandey},
the stretched time $\xi=\omega t$ and the slow time $\eta=\epsilon t$
and expand the forcing frequency $\omega$ about the natural frequency
of the oscillator i.e. 1 with $\epsilon$ as the expansion parameter,\begin{equation}
\omega=1+k\epsilon+{\cal O}(\epsilon^{2}),\label{eq:w_expansion}\end{equation}
where $k$ is a detuning parameter at order $\epsilon$. The variable
$x$ now is expanded in a power series\begin{equation}
x=x_{0}(\xi,\eta)+\epsilon x_{1}(\xi,\eta)+{\cal O}(\epsilon^{2}).\label{eq:x_expansion}\end{equation}
Substituting Eqs.(\ref{eq:w_expansion}) and (\ref{eq:x_expansion})
in Eq.(\ref{eq:entrain}) and collecting terms at the order $\epsilon=0$
and 1, we have,\begin{eqnarray}
x_{0\xi\xi}+x_{0} & = & 0,\label{eq:first_order}\\
 &  & ,\label{eq:second_order}\end{eqnarray}
where the subscripts refer to derivatives with respect to $\xi$ and
$\eta$. The solution to Eq.(\ref{eq:first_order}) can be taken as\begin{equation}
x_{0}(\xi,\eta)=A(\eta)\,\cos\xi+B(\eta)\,\sin\xi,\label{eq:solution1}\end{equation}
where the coefficients $A$ and $B$ are functions of the slow time-scale.
Substituting Eq.(\ref{eq:solution1}) into Eq.(\ref{eq:second_order})
and removing the secular terms \cite{strogatz}, we have the following
coupled differential equations for the slow time-scale,\begin{eqnarray}
A' & = & -kB+\frac{1}{2}\mu A-\frac{1}{8}\mu A(A^{2}+B^{2})+\frac{1}{4}\lambda B,\label{eq:A}\\
B' & = & kA+\frac{1}{2}\mu B-\frac{1}{8}\mu B(A^{2}+B^{2})+\frac{1}{4}\lambda A.\label{eq:B}\end{eqnarray}
We note that the hyperbolic fixed points of the slow flow correspond
to the periodic motion of the original equation Eq.(\ref{eq:entrain})
i.e. an entrainment and limit cycles of the slow flow correspond to
quasi-periodic motions of Eq.(\ref{eq:entrain}) \cite{guckenheimer}.
From simple resonance dynamics it is evident that the entrainment
region of Eq.(\ref{eq:entrain}) by the parametric forcing term should
increase as the parametric forcing amplitude $\lambda$ increases,
allowing an wider range in the detuning parameter $k$ during which
the entrainment is observed. Therefore it is worthwhile to study the
flow of the slow variables through Eqs.(\ref{eq:A}) and (\ref{eq:B})
as we vary the parameters $k$ and $\lambda$.%
\begin{figure}
\begin{centering}
\includegraphics[width=0.5\textwidth]{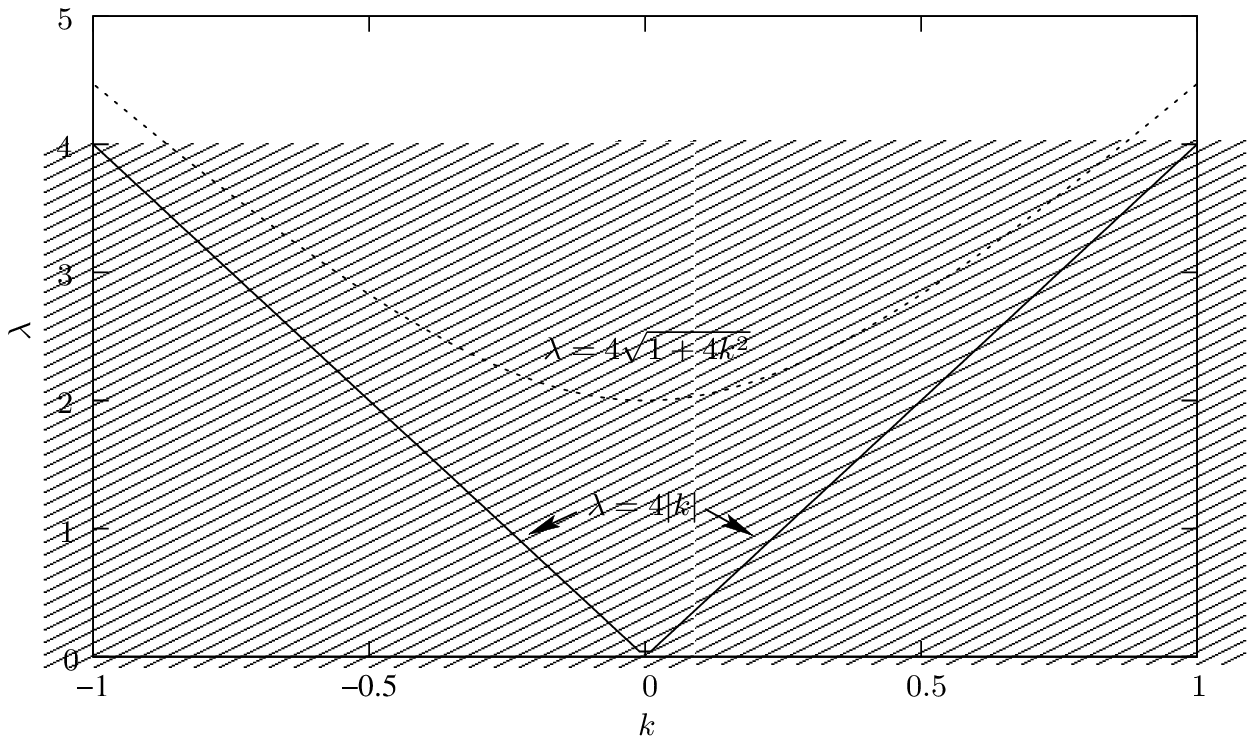}
\par\end{centering}

\caption{\label{fig:linear-bifurcation}Bifurcation diagram of Eqs.(\ref{eq:A})
and (\ref{eq:B}) in the $k$-$\lambda$ plane. The line $\lambda=4|k|$
along which a Hopf bifurcation occurs at the origin. Below this line,
in the shaded region, there exists a limit cycle. The dashed line
denote a saddle-node bifurcation at the origin.}
\end{figure}

\begin{figure}
\begin{centering}
\includegraphics[width=0.5\textwidth]{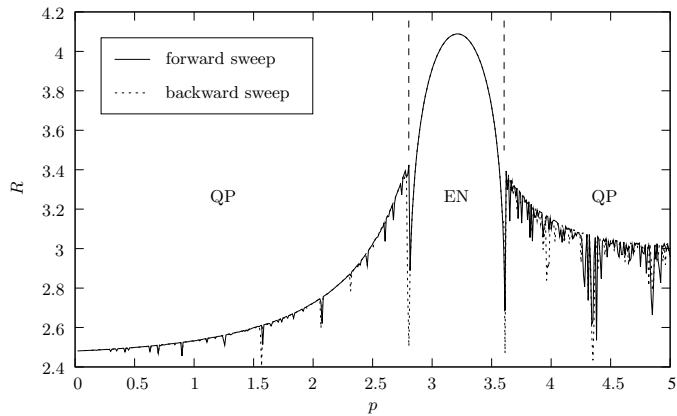}
\par\end{centering}

\caption{\label{fig:entrainment}Entrainment by the parametric dust-charge
fluctuation for $\alpha=0.15,\beta=0.1,\lambda=1.0,\omega_{d}=1.0,\epsilon=0.5$,
where the amplitude $R$ of the oscillation is plotted against the
period $p$ of the parametric forcing term. The two vertical dashed
lines at $p=2.8$ and $3.6$ indicate the region of entrainment (marked
as EN). The other behavior is quasi-periodic (marked by QP). Only
very little hysteresis is observed in the quasi-periodic regions.}
\end{figure}

In Fig.\ref{fig:linear-bifurcation}, the bifurcation diagram of the
slow flow in the $k$-$\lambda$ plane is shown. Along the line $\lambda=4|k|$,
a Hopf bifurcation occurs at the origin, below which, in the shaded
region, a limit cycle appears. As $\lambda$ falls below $4|k|$,
the fixed point at the origin becomes an unstable spiral {[}complex
eigenvalues of the linearized Jacobian of Eqs.(\ref{eq:A}) and (\ref{eq:B})]
from an unstable node. Along the line $\lambda=4\sqrt{1+4k^{2}}$,
saddle node bifurcation occurs for sufficiently high $\lambda$ when
the origin becomes a saddle point from an unstable node after two
other saddle points coalesce at the origin. From what we have observed,
one can conclude that entrainment by parametric forcing occurs above
the shaded region of Fig.\ref{fig:linear-bifurcation}. In the shaded
region, we expect a quasi-periodic behavior. These results are confirmed
from the numerical solutions of Eq.(\ref{eq:entrain}), which are
shown in Fig.\ref{fig:entrainment}. The entrainment region for $\lambda=1$
can be calculated from Fig.\ref{fig:linear-bifurcation} as $-0.25<k<+0.25$,
which in terms of period of Eq.(\ref{eq:entrain}) for the parameters
of Fig.\ref{fig:linear-bifurcation}, is given by $2.79<p<3.59$ which
very well agrees with the numerical results (see Fig.\ref{fig:entrainment}).

As we increase $\epsilon$, the expansion parameter, the prediction
from Fig.\ref{fig:linear-bifurcation} however agrees less and less
as the system makes a transition to chaos.

\section{Stability and bifurcations}

The stability of a periodic orbit can be effectively studied using
Floquet theory \cite{guckenheimer,kocak}. In this section, we briefly
review the Floquet theory with reference to Eq.(\ref{eq:dusteq}).
We first write the second order dynamical equation Eq.(\ref{eq:dusteq})
as two first order equations,\begin{eqnarray}
\dot{x} & = & y,\label{eq:first_order1}\\
\dot{y} & = & (\alpha-\beta x^{2})y-\omega_{d}^{2}x(1-\epsilon\lambda\,\cos\nu t).\label{eq:first_order2}\end{eqnarray}
We have already shown in the previous section that a frequency-locked
(entrained) phase with a single periodic limit cycle can exist for
Eqs.(\ref{eq:first_order1},\ref{eq:first_order2}). In this section,
we are going to study the stability of this periodic orbit and bifurcations
leading to the onset of chaos.

The Poincar\'e map of an initial point $z_{0}=(x_{0},y_{0})$ on
the periodic orbit (limit cycle) can be obtained by sampling the orbit
points $z_{n}$ at discrete time interval $t=t_{n},\, n=1,2,3,\ldots$.
So the transformation which successively maps the Poincar\'e section
$P(z)$ is $z_{n+1}=P(z_{n})$. The linear stability of a $q$-periodic
orbit with $P^{q}(z_{0})=z_{0}$ can now be determined from the linearized
map given by the matrix $DP^{q}$ of $P^{q}$ at an orbit point $z_{0}$,
where $P^{q}$ is the $q$-times iterated Poincar\'e map. The linearized
matrix $M=DP^{q}$ can be obtained by integrating the linearized equations
corresponding to Eqs.(\ref{eq:first_order1},\ref{eq:first_order2})
for small perturbations along the $q$-periodic orbit \cite{guckenheimer}.

Assume that $z^{\star}(t)=z^{\star}(t+q)$ is a point lying on the
$q$-periodic limit cycle of Eqs.(\ref{eq:first_order1},\ref{eq:first_order2}).
We perturb the orbit with a small perturbation $\delta z=(\delta x,\delta y)$
and linearize Eqs.(\ref{eq:first_order1},\ref{eq:first_order2})
about the closed orbit,\begin{equation}
\left(\begin{array}{c}
\dot{\delta x}\\
\dot{\delta y}\end{array}\right)=J(t)\left(\begin{array}{c}
\delta x\\
\delta y\end{array}\right),\quad J(t)=\left(\begin{array}{cc}
0 & 1\\
f_{x} & f_{y}\end{array}\right)_{(x,y)=(x^{\star},y^{\star})},\label{eq:jcobian}\end{equation}
where $J(t)$ is the $q$-periodic linearized Jacobian. The partial
derivatives $f_{x,y}$ are given by,\begin{equation}
f_{x}=-2\beta xy-\omega_{d}^{2}(1-\epsilon\lambda\,\cos\nu t),\quad f_{y}=(\alpha-\beta x^{2}).\label{eq:fxy}\end{equation}
We now assume that $W(t)=[w_{1}(t),w_{2}(t)]$ is a fundamental solution
matrix with $W(0)=I$ \cite{guckenheimer}. The general solution of
the $q$-periodic system, Eq.(\ref{eq:jcobian}), is then given by,\begin{equation}
\left(\begin{array}{c}
\delta x(t)\\
\delta y(t)\end{array}\right)=W(t)\left(\begin{array}{c}
\delta x(0)\\
\delta y(0)\end{array}\right).\label{eq:fundamental}\end{equation}
We then substitute Eq.(\ref{eq:fundamental}) into Eq.(\ref{eq:jcobian})
to obtain the initial value problem,\begin{equation}
\dot{W}(t)=J(t)W(t),\quad W(0)=I,\label{eq:initial}\end{equation}
where $W(q)$ is the linearized map $DP^{q}(z_{0})$. So, the matrix
$DP^{q}$ can, in principle, be obtained from numerical integration
of Eq.(\ref{eq:initial}) over period $q$. However, the numerical
procedure is not very straight forward and requires sophisticated
techniques to determine the exact form of the matrix $DP^{q}$ which
is very sensitive to initial conditions $(x^{\star},y^{\star})$.
The matrix $DP^{q}$ is known as the \emph{monodromy} matrix for the
$q$-periodic orbit and the eigenvalues of this monodromy matrix,
popularly known as the Floquet multipliers \cite{guckenheimer,kocak},
indicate the stability of the $q$-periodic orbit. Therefore, the
values of the Floquet multipliers have to be determined with considerable
precision for understanding the true nature of the stability of the
nonlinear system. The characteristic equation of the linearized map
$M=DP^{q}$ is given by,\begin{equation}
\zeta^{2}-\tau\zeta+\Delta=0,\label{eq:charateristic}\end{equation}
where the eigenvalues $\zeta_{1,2}$ are the Floquet multipliers and
$\tau={\rm tr}(M),\Delta={\rm det}(M)$. The determinant $\Delta$
is given by \cite{arnold,guckenheimer,kocak}\begin{eqnarray}
\Delta & = & e^{\int_{0}^{q}\,{\rm tr}(J)\, dt}=e^{(\alpha-\beta x^{\star2})q},\label{eq:floquet_det}\\
\tau & = & \frac{1}{q}\int_{0}^{q}\,{\rm tr}(J)\, dt\,\left({\rm mod}\frac{2\pi i}{q}\right).\label{eq:floquet_tr}\end{eqnarray}

\begin{figure}
\begin{centering}
\includegraphics[width=0.5\textwidth]{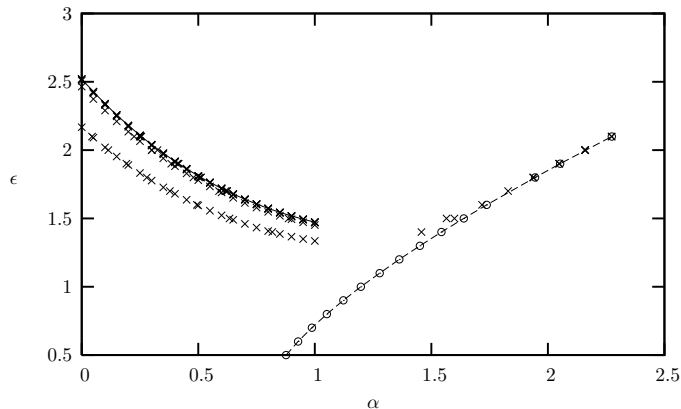}
\par\end{centering}

\caption{\label{fig:scaling}Stability diagram of Eqs.(\ref{eq:first_order1},\ref{eq:first_order2})
in the $\alpha$-$\epsilon$ plane. The period doubling bifurcations
occur along the lines denoted by the points `$\times$'. The points
denoted by a `$\circ$' along the dashed line indicate pitchfork bifurcations.
The chaotic regime lies above the two lines. As we can see that some
pitchfork bifurcations are followed by a period doubling cascade.
The other parameters are $\beta=0.1$ and $\nu=\omega_{d}=\lambda=1.0$.}
\end{figure}

We know from Floquet theory that the periodic orbit is stable only
if the pair of Floquet multipliers lie inside the unit circle. The
bifurcations of the periodic orbit occur on the unit circle. From
the expression for $\Delta$, Eq.(\ref{eq:floquet_det}), it can be
seen that we do have bifurcations depending on a balancing of the
parameters $\alpha,\beta$, and the periodic orbit, which is determined
by the parameter $\epsilon$, the magnitude of the parametric driving
force. In all probability, the unstable region should lie in the region
of large $\alpha$. We numerically determine the Floquet multipliers
for a range of periodic orbit in the parameter space $(\alpha,\epsilon)$
with $\beta=0.1$ and $\omega_{d}=\lambda=1.0$ and the resultant
stability diagram is shown in Fig.\ref{fig:scaling}. In Fig.\ref{fig:scaling},
period doubling bifurcations occur along the lines denoted by a `$\times$'
sign and subcritical pitchfork bifurcations occur along the dashed
line denoted by a `$\circ$'. The line joining the `$\times$' points
in the figure denotes the accumulation points or the limiting points
of the period doubling bifurcations before transition to chaos. The
chaotic regime lies above these lines. As the two lines seem to intersect,
when extended, we see that some of the period doubling cascades are
preceded by pitchfork bifurcations. %

\begin{figure}
\begin{centering}
\includegraphics[width=0.49\textwidth]{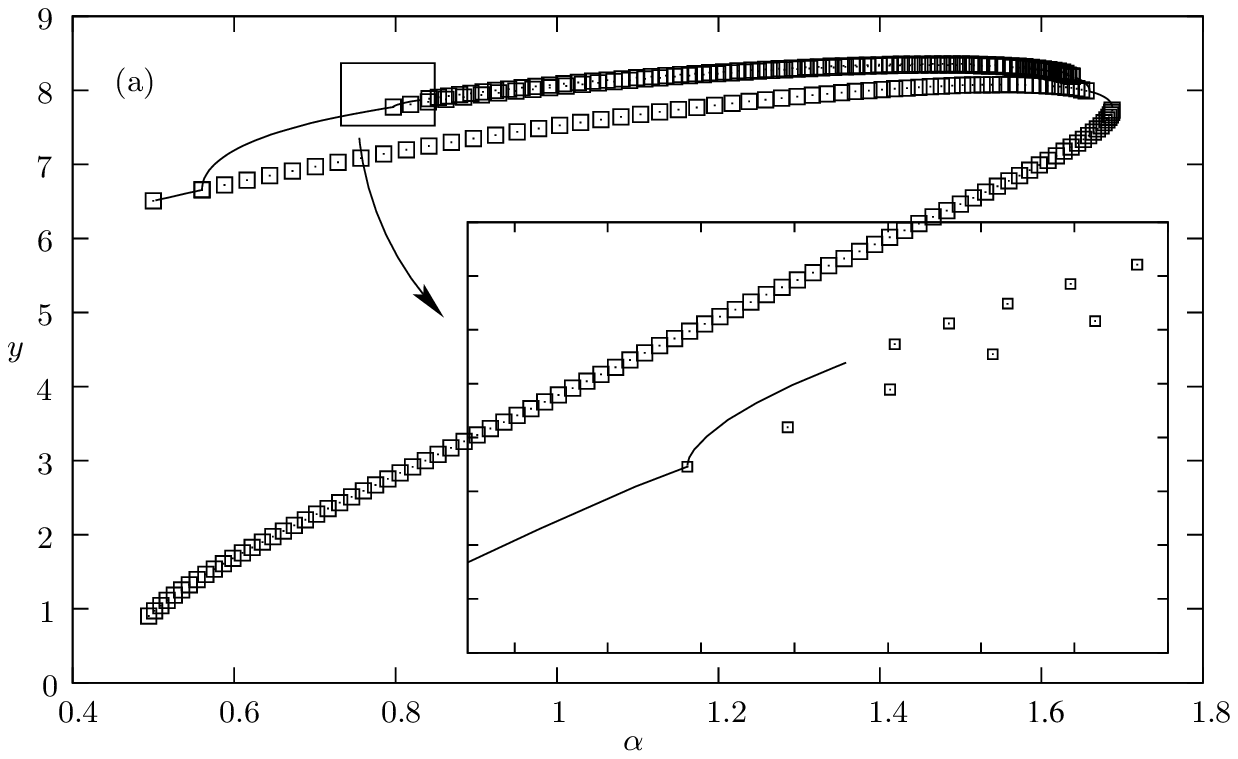}\hfill{}\includegraphics[width=0.49\textwidth]{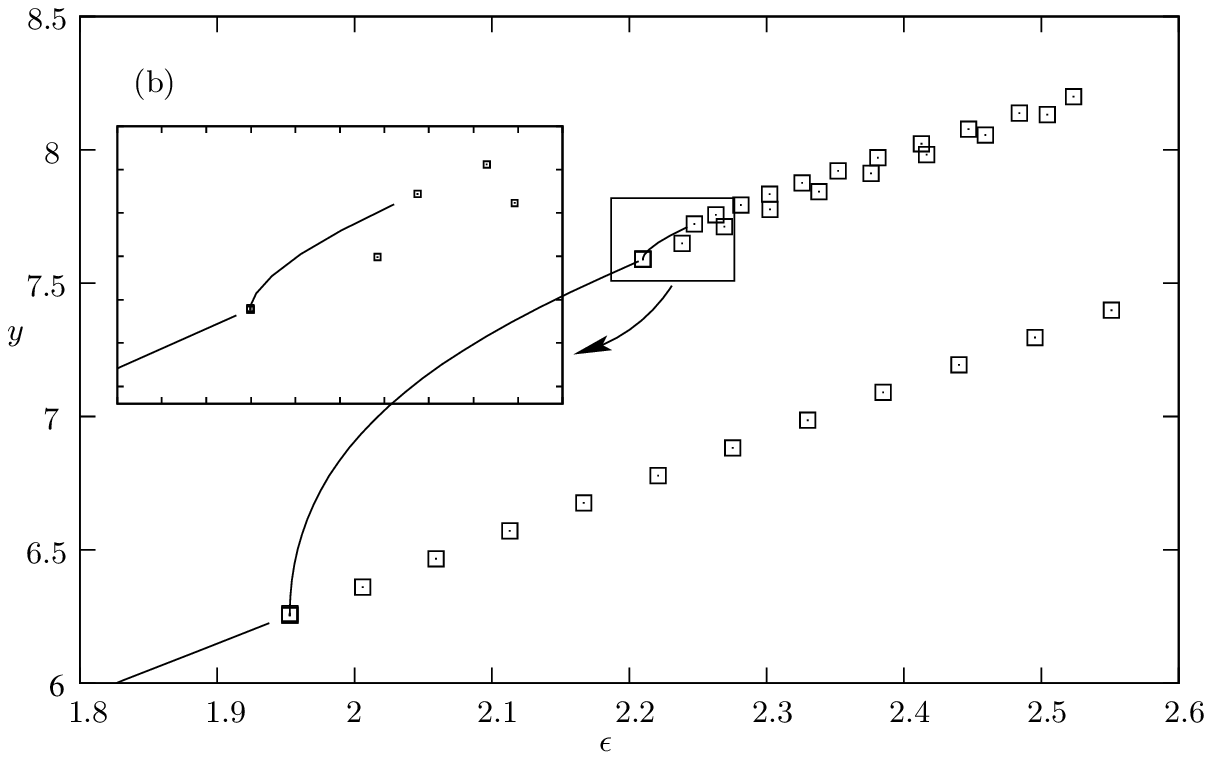}
\par\end{centering}

\caption{\label{fig:bifurcation}Bifurcation diagrams for the period doubling
sequences. In (a), the bifurcation parameter is $\alpha$ with $\epsilon=1.5$
and in (b), the bifurcation parameter is $\epsilon$ with $\alpha=0.15$.
All other parameters are same as in Fig.\ref{fig:scaling}. The \emph{open
boxes} in the diagrams are unstable orbits and the solid lines indicate
stable periodic orbits (limit cycles) with period doubling bifurcations.
Blow-up regions of the second period double cascade is shown in the
insets.}
\end{figure}

As the pair of Floquet multipliers decreases through $-1$ at the
period doubling points, the $q$-periodic orbit loses its stability
to jump to a $2q$-periodic stable limit cycle. So, in the observed
parameter regime of Fig.\ref{fig:scaling}, all period doubling bifurcations
are supercritical \cite{strogatz}. In case of the pitchfork bifurcations,
the Floquet multipliers increases through $+1$ and are subcritical
as there are no stable limit cycles after the bifurcations and the
system becomes aperiodic \cite{strogatz}. In Figs.\ref{fig:scaling}(a)
and (b), two successive period doubling orbits are shown. In Figs.\ref{fig:bifurcation}(a)
and (b), we have shown the bifurcation diagrams with the bifurcation
parameters as $\alpha$ and $\epsilon$ (for details, please see the
captions in the figure). These bifurcation diagrams are obtained with
the help of AUTO \cite{doedel} as a part of the XPPAUT package \cite{ermentrout}.
\begin{table}

\caption{\label{tab:scaling}Scaling of period doubling cascades.}

\begin{centering}
\begin{tabular}{cccccccccccccc}
\hline 
$k$&
&
$\alpha_{k}$&
&
$\delta_{k}$&
&
&
&
&
&
&
$\epsilon_{k}$&
&
$\delta_{k}$\tabularnewline
\hline
\hline 
1&
&
0.3781420&
&
&
&
&
&
&
&
&
1.952855&
&
\tabularnewline
2&
&
0.5888546&
&
7.61&
&
&
&
&
&
&
2.209903&
&
6.85\tabularnewline
3&
&
0.6165303&
&
4.77&
&
&
&
&
&
&
2.247453&
&
4.73\tabularnewline
4&
&
0.6223341&
&
4.70&
&
&
&
&
&
&
2.255401&
&
4.70\tabularnewline
5&
&
0.6235698&
&
4.67&
&
&
&
&
&
&
2.257093&
&
4.65\tabularnewline
6&
&
0.6238345&
&
&
&
&
&
&
&
&
2.257457&
&
\tabularnewline
\hline
\end{tabular}
\par\end{centering}
\end{table}

\subsection{Scaling of the period doubling cascades}

It is interesting to investigate the scaling behaviour of the period
doubling cascades in light of the scaling of the period doubling sequences
in 1-D maps. As usually observed in any period doubling cascades,
the bifurcation parameter, which in our case are $\alpha$ and $\epsilon$,
converges geometrically to a limiting value with the convergence ratio
approaching a unique value, analogous to Feigenbaum number in case
of 1-D maps \cite{strogatz,feigenbaum}. In Table.\ref{tab:scaling},
we have listed the values of the bifurcation parameters as these converge
to their respective limiting values. The ratio of this convergence
$\delta_{k}$, expressed as,\begin{equation}
\delta_{k}=\frac{A_{k}-A_{k-1}}{A_{k+1}-A_{k}},\qquad\lim_{k\rightarrow\infty}\delta_{k}=\delta,\label{eq:delta_k}\end{equation}
approaches a unique limiting value $\delta$ as the bifurcation parameter
$A_{k}\rightarrow{\rm const.}$, with $k$ denoting each successive
period doubling point. The value of $\delta$ characterizes the scaling
property of the bifurcation, which agrees well with the Feigenbaum
number $4.669\ldots$ for 1-D maps \cite{feigenbaum,jeong,kim}. %
\begin{figure}
\begin{centering}
\includegraphics[width=0.49\textwidth]{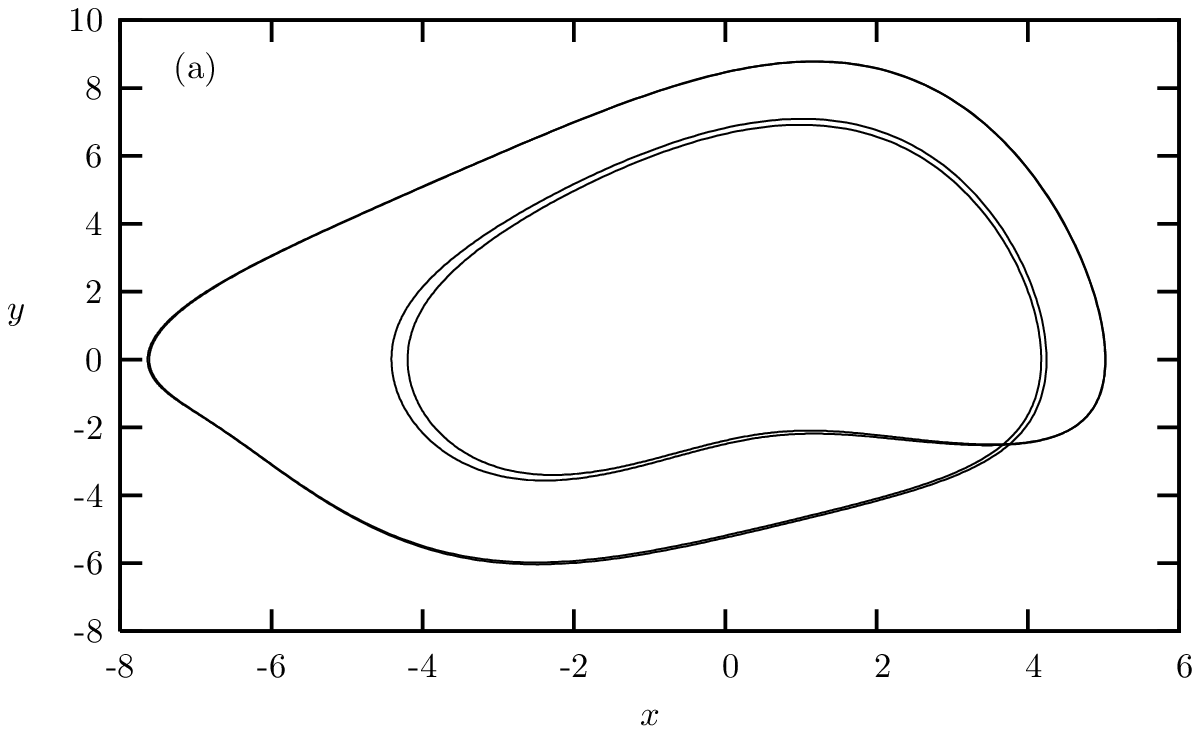}\hfill{}\includegraphics[width=0.49\textwidth]{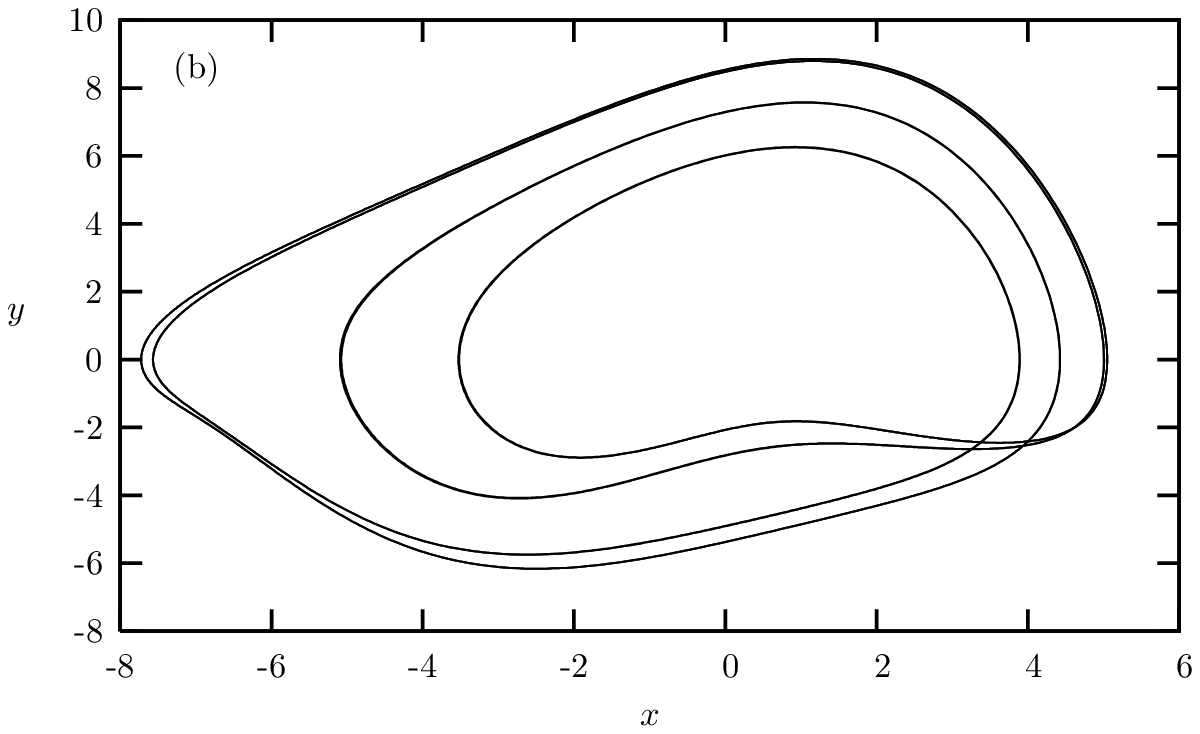}
\par\end{centering}

\caption{\label{fig:doubling}Phase portrait of two successive period doubling
orbits for $\alpha=0.5$ and (a) $\epsilon=1.779955$ and (b) $\epsilon=1.805263$.
Rest of the parameters are same as in Fig.\ref{fig:scaling}.}
\end{figure}

\section{Transition to chaos}

In this section, we study the chaotic behavior of Eq.(\ref{eq:dusteq}).
The single most prominent feature of chaos is its sensitive dependence
on initial conditions, which is measured by the Lyapunov characteristic
exponents (LCEs) \cite{shimada}. These exponents are invariant global
indicators of the non-linear system. For a continuous dynamical system
described by a set of autonomous ordinary differential equations,
the number of LCEs is equal to the dimension of the system. By discretizing
the temporal dimension as $\Delta t$, the LCEs can be defined as,\begin{equation}
\sigma_{i}=\lim_{N\rightarrow\infty}\lim_{\Delta t\rightarrow0}\frac{1}{N\,\Delta t}\,\ln[S_{i}(M_{N})],\label{eq:lce}\end{equation}
where\begin{equation}
M_{N}=\prod_{i=0}^{N}e^{J(i\Delta t)\Delta t}.\label{eq:mn}\end{equation}
In the above relations, $S_{i}$ are the singular values of the matrix
$M_{N}$ and $J$ is the Jacobian. Numerically, the number $N$ denotes
the number of integration steps of length $\Delta t$. Here, we employ
a numerical algorithm, based on the Wolf's well known method to calculate
the LCEs \cite{wolf}. Note that, Wolf's original method is not the
best approach to calculate the LCEs and it can be modified to contain
the non-uniformity-factors of the LCEs \cite{grond}. An important
necessary step in Wolf's algorithm is to re-orthogonalize of the set
of vectors, which is carried out, usually, through Gram-Schmid orthogonalization
procedure. In our numerical routine too, we have used the Gram-Schmid
orthogonalization to calculate the LCEs. In particular, we have calculated
the LCEs for the following autonomous system,\begin{equation}
\begin{array}{rcl}
\dot{x} & = & y,\\
\dot{y} & = & (\alpha-\beta x^{2})y-\omega_{d}^{2}x(1-\epsilon\lambda\, u),\\
\dot{u} & = & u(1-u^{2}-v^{2})-2\pi v/T,\\
\dot{v} & = & v(1-u^{2}-v^{2})+2\pi u/T,\end{array}\label{eq:auto}\end{equation}
where $\nu=2\pi/T$. Note that, we have made the original non-autonomous
equation, Eq.(\ref{eq:dusteq}), a system of autonomous equations
by introducing the last two equations of Eqs.(\ref{eq:auto}), which
have stable and unique solutions,\begin{equation}
u(t)=\cos(2\pi t/T),\quad v(t)=\sin(2\pi t/T),\label{eq:auto_soln}\end{equation}
for the initial values $[u(0),v(0)]=[1,0]$.

\begin{figure}
\begin{centering}
\includegraphics[width=0.49\textwidth]{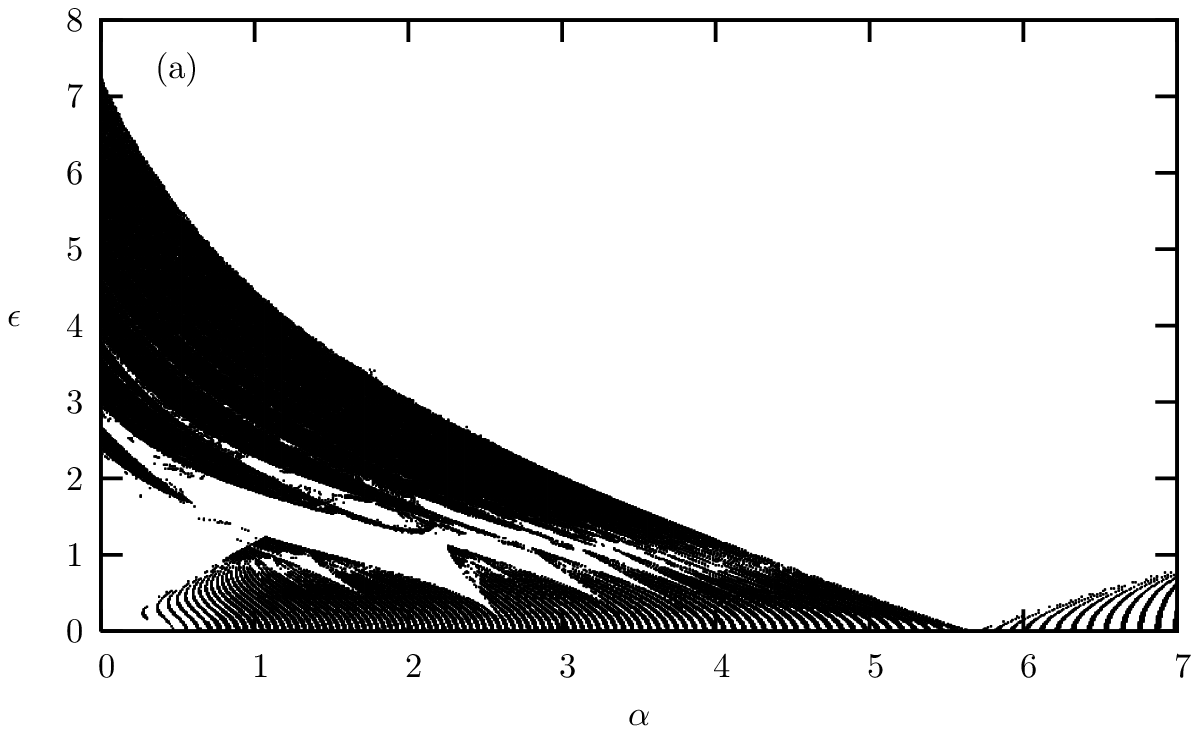}\hfill{}\includegraphics[width=0.49\textwidth]{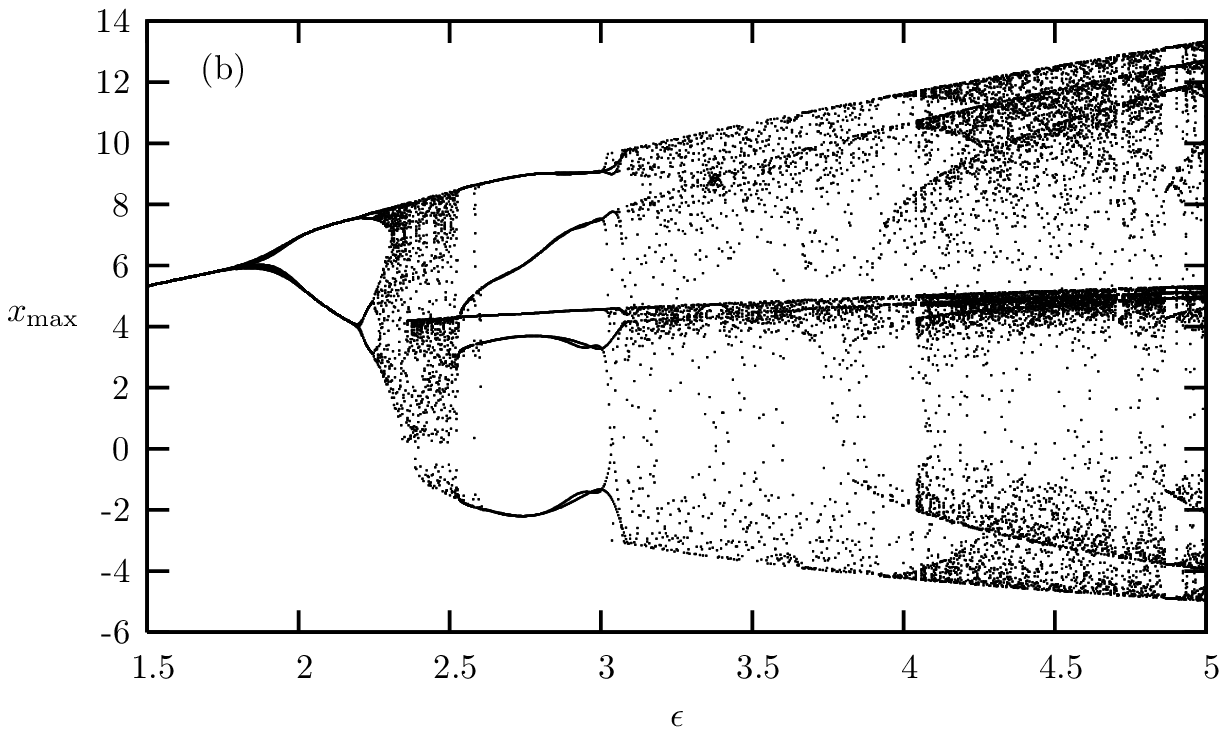}
\par\end{centering}

\caption{\label{fig:lyapunov-orbit}(a) Plot of the maximal Lyapunov exponent,
$\sigma_{1}$, for Eq.(\ref{eq:dusteq}) in the parameter space of
$\alpha$-$\epsilon$. The Lyapunov exponent is calculated in the
entire parameter space indicated in the figure. The \emph{blackened}
portions indicate a positive and the blank portions of the figure
indicate a negative Lyapunov exponent. As can be seen from the figure,
the fractal nature of the chaos is indicated by repetitive appearance
of the whole figure in the smaller regions close to the horizontal
axis. It also seems that the entire figure is \emph{only} a part of
a large figure covering the entire $\alpha$-$\epsilon$ plane. (b)
The orbit diagram for Eq.(\ref{eq:auto}), where the successive local
maxima are plotted against the bifurcation parameter, for $\alpha=0.15.$
Other parameters are same as in (a).}
\end{figure}

\begin{figure}
\begin{centering}
\includegraphics[width=0.49\textwidth]{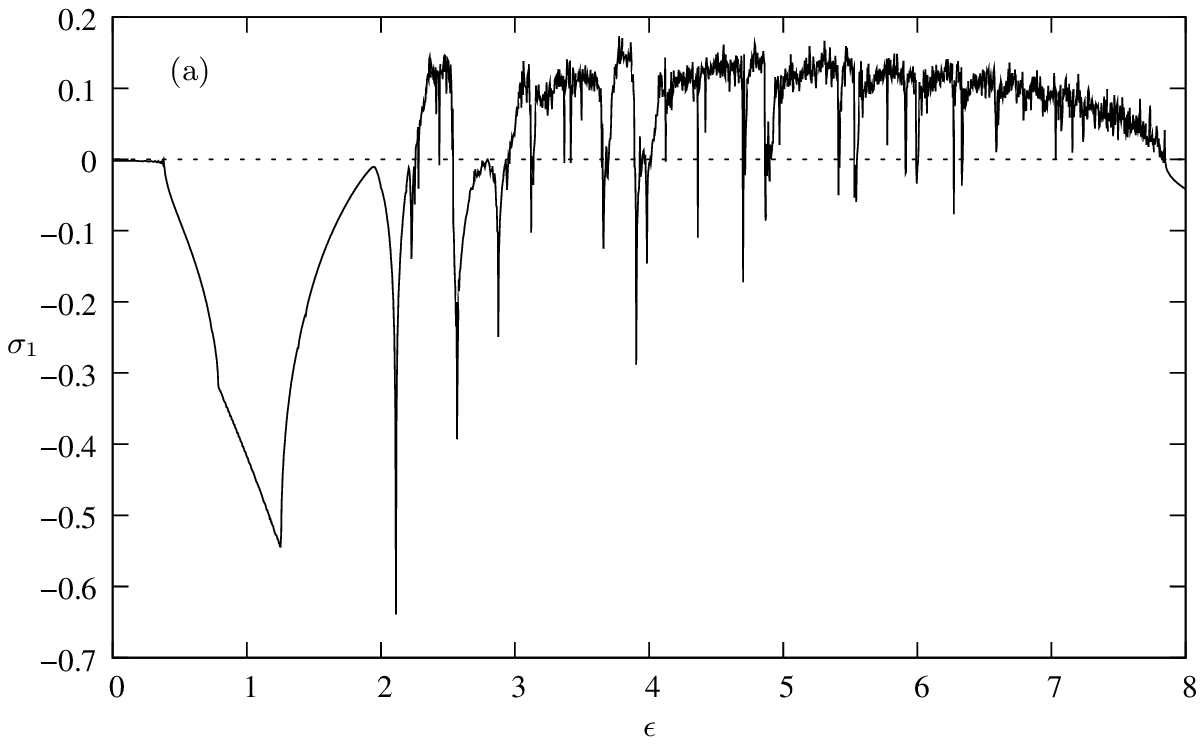}\hfill{}\includegraphics[width=0.49\textwidth]{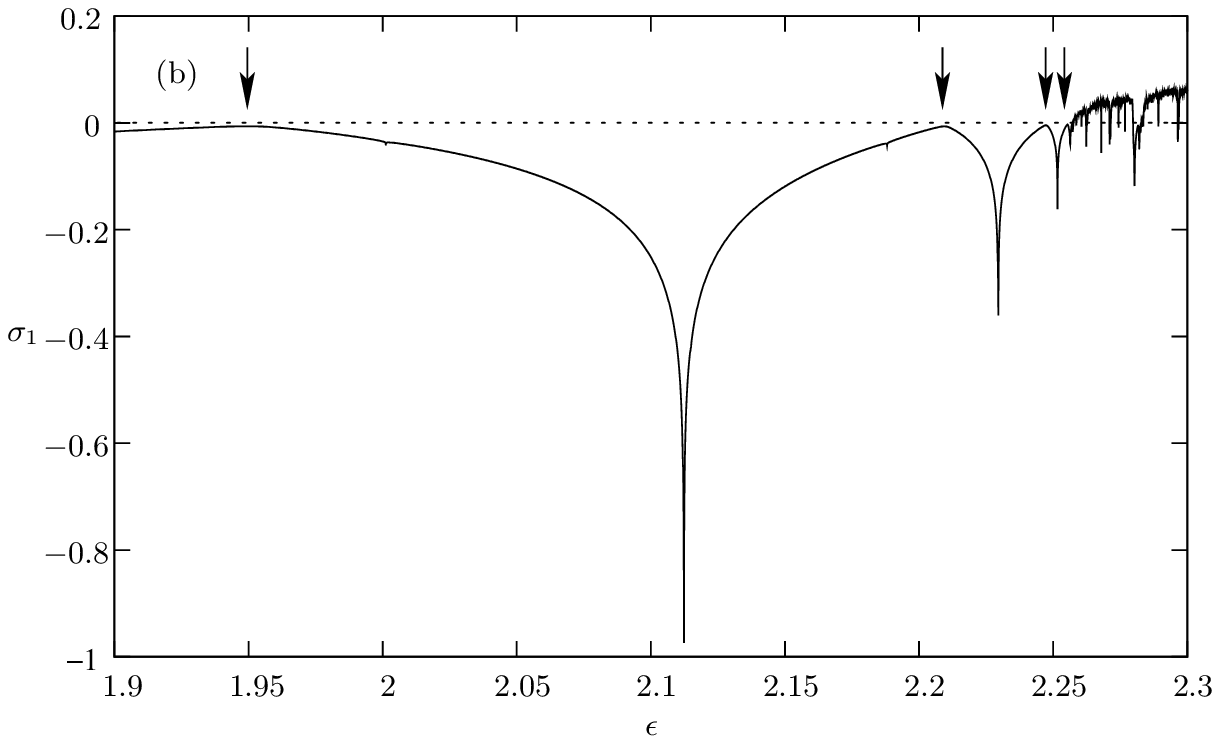}\\
~\\
\includegraphics[width=0.49\textwidth]{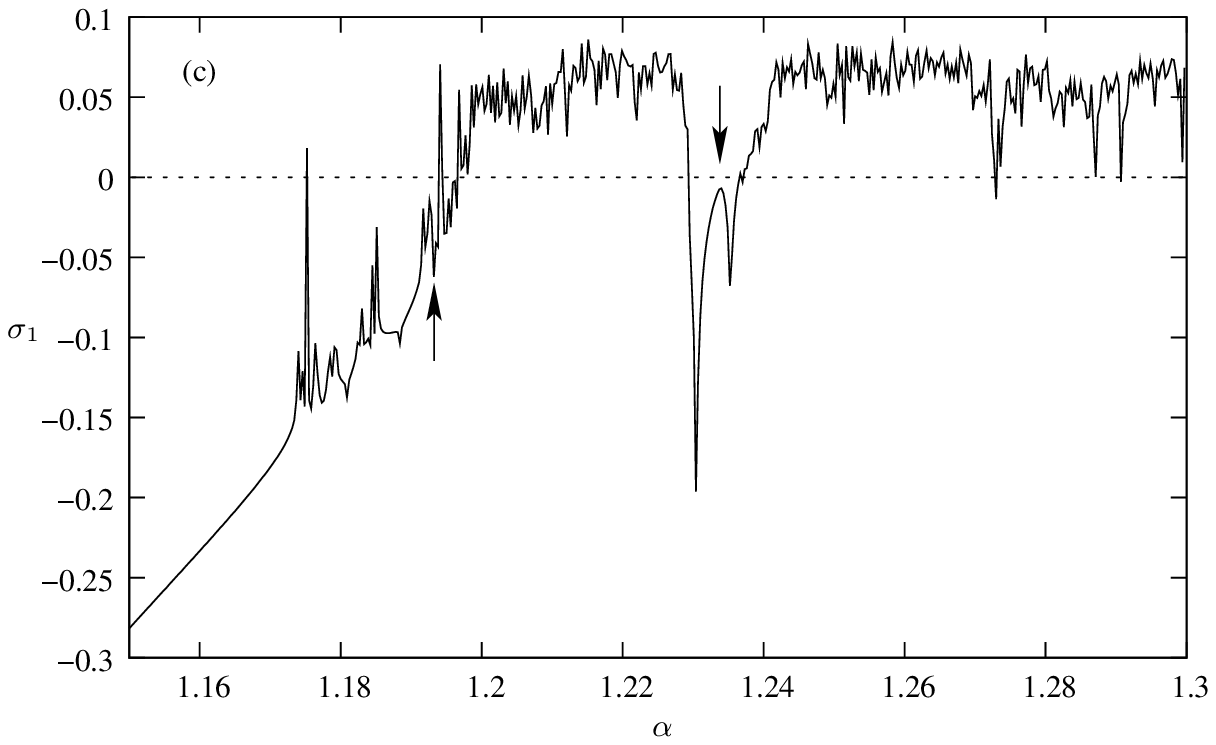}
\par\end{centering}

\caption{\label{fig:lyapunov_curves}(a) Plot of the maximal Lyapunov exponent
$\sigma_{1}$ against the bifurcation parameter $\epsilon$ for $\alpha=0.15$.
Other parameters are same is in Fig.\ref{fig:lyapunov-orbit}. (b)
A blow up of the previous figure is shown where the period doubling
points are marked with arrows. Note that after each period doubling
bifurcation, the system transits to a stable limit cycle (characterized
by large negative $\sigma_{1}$ before getting unstable for the next
period doubling point when $\sigma_{1}$ becomes close to zero. (c)
The Maximal LCE in a region of pitchfork bifurcation. The first arrow
(up) indicates the pitchfork bifurcation after which the system transits
to a chaotic regime before becoming periodic (denoted by the negative
LCE) and becomes chaotic again as $\alpha$ increases, following a
period doubling cascade, denoted by the second arrow (down).}
\end{figure}

In Fig.\ref{fig:lyapunov-orbit}(a), we have plotted the maximal LCE
for the Eqs.(\ref{eq:auto}), in the parameter space of $\alpha$-$\epsilon$
for the range shown. The parameters are $\beta=0.1,\omega_{d}^{2}=1$,
and $\lambda=1$. The period of the orbit is chosen as $T=6.5$ which
corresponds to $\nu=0.96664$. The \emph{blackened} portions of the
figure indicate a positive Lyapunov exponent indicating chaos. In
all other places the maximal Lyapunov exponent is negative, signifying
stable oscillations of the system. We can see the fractal behavior
of the chaos from the figure. It also seems that the whole figure
is only a part of a larger figure covering the entire domain of the
$\alpha$-$\epsilon$ plane. In Fig.\ref{fig:lyapunov_curves}, the
maximal LCE, $\sigma_{1}$, is plotted along with a blow up of the
region of period doubling cascades. The period doubling points are
marked with `arrows' in Fig.\ref{fig:lyapunov_curves}(b).

\begin{figure}
\begin{centering}
\includegraphics[width=0.49\textwidth]{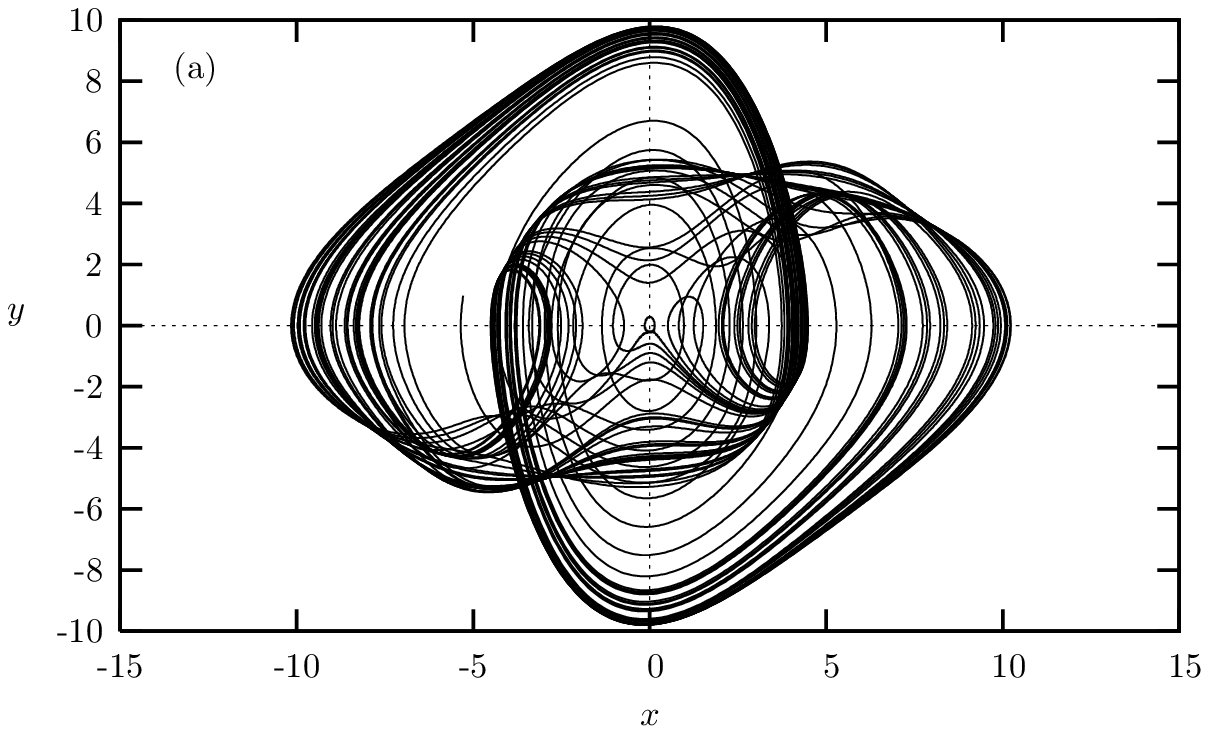}\hfill{}\includegraphics[width=0.493\textwidth]{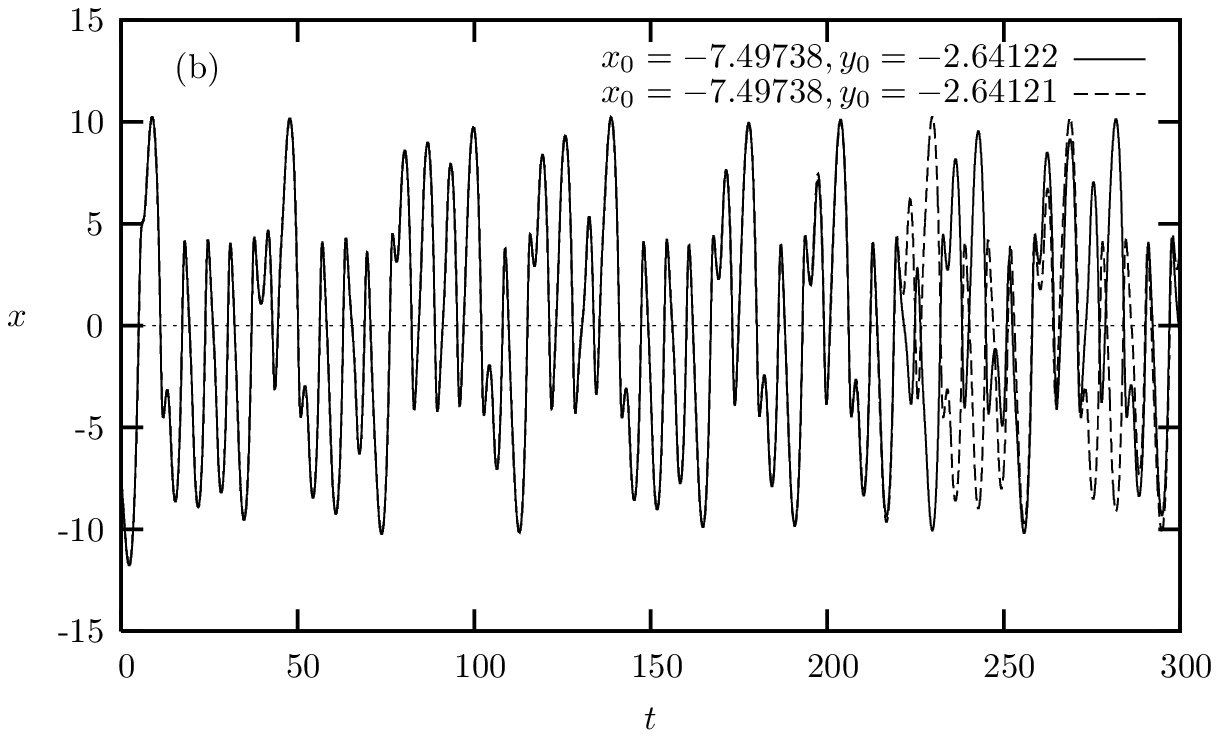}\\
~\\

\par\end{centering}

\begin{centering}
\includegraphics[width=0.455\textwidth]{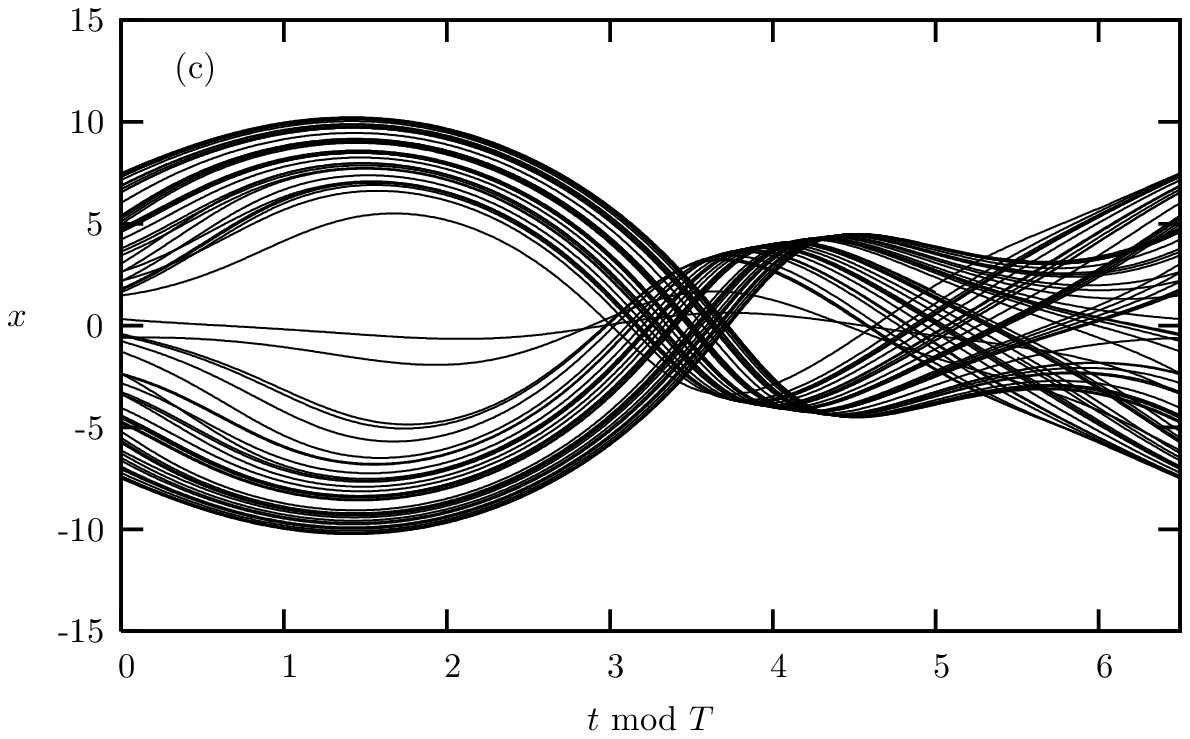}\hfill{}\includegraphics[width=0.49\textwidth]{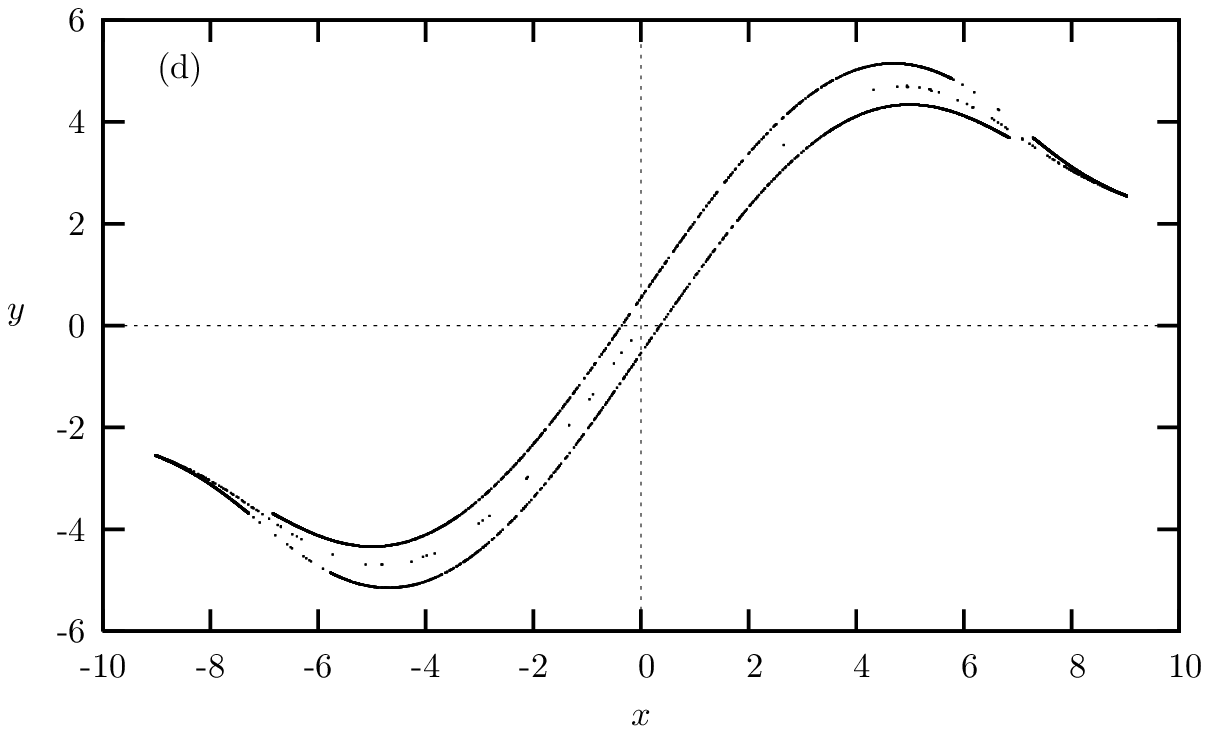}
\par\end{centering}

\caption{\label{fig:chaos}(a) Projection of the phase portrait in the $x$-$y$
plane in the chaotic regime of Eq.(\ref{eq:dusteq}). (b) The corresponding
time evolution of the variable $x$. Note the divergence of the two
curves for change in the initial condition $(x_{0},y_{0})$ at $t=0$
at the fifth decimal place, showing the sensitivity of the system
on initial conditions in the chaotic region. (c) The phase portrait
in the cylindrical space $x\times y\times(t\,{\rm mod}\, T)$, where
$T$ is the fundamental period of the system, which is $6.5$ in this
case. (d) The corresponding Poincar\'e map, which seems to fall in
a fractal set, signifying chaos. The parameters in all the figures
are chosen for the chaotic regime, $\alpha=0.07,\epsilon=3.41,\lambda=1,\omega_{d}=1$,
and $\nu=0.96664$ corresponding to $T=6.5$ (refer to Fig.\ref{fig:lyapunov-orbit}).}
\end{figure}

We have constructed an \emph{orbit} diagram \cite{strogatz} {[}Fig.\ref{fig:lyapunov-orbit}(b)]
for the system, Eq.(\ref{eq:auto}), where we have plotted the successive
local maxima for the variable $x$ of the oscillation against the
bifurcation parameter $\epsilon$. The period doubling bifurcations
occur near $1.9$ and $2.2$ before the system becomes chaotic. Note
the appearance of 5-period window in the region $\epsilon=2.5$ and
$3.0$ after which the system becomes chaotic again. The chaotic orbit
along with the time evolution of the system in the chaotic regime
and the Poincar\'e map are shown in Fig.\ref{fig:chaos}.


\section{Conclusion}

In this work, we have carried out a detailed investigation of the
stability, bifurcation leading to chaos for the vdPM system arising
out dust-charge fluctuation in a a dusty plasma. We have shown that
the system can be highly chaotic depending the chosen parameters and
not as restrictive as has been pointed out by Saitou and Honzawa \cite{saitou}.
In fact, an wide range of chaotic region exists for the parametric
driving strength ($\epsilon\lambda$) as low as 0.05, as shown by
the plot of the maximal Lyapunov exponent $\sigma_{1}$. It has also
been found that the system can be completely deterministic in the
middle of two chaotic regions and exhibit quasi-periodicity, as shown
by the orbit diagram where a 5-period window appears in the middle
of a chaotic region. We have further shown that when the parametric
forcing term related to dust-charge fluctuation is small, away from
the chaotic regime, the system can be driven in a frequency-locked
state when a harmonic resonance of 2:1 takes place between the driving
frequency and the fundamental frequency of the system. We have found
that in most of the cases, the system transits to chaos through a
cascade of period doubling bifurcations and the scaling of the period
doubling cascades closely agrees to that of 1-D maps \cite{feigenbaum}.

\end{document}